\documentclass{article}
\usepackage{spconf,amsmath,graphicx, amsfonts}
\usepackage{makecell}
\usepackage{multirow}
\usepackage{booktabs}
\usepackage{rotating}
\usepackage{enumitem}
\usepackage{cite}
\usepackage[colorlinks,
            linkcolor=green,       %
            anchorcolor=blue,  %
            urlcolor=blue,        %
            ]{hyperref} 

\makeatletter
\def\UrlAlphabet{%
      \do\a\do\b\do\c\do\d\do\e\do\f\do\g\do\h\do\i\do\j%
      \do\k\do\l\do\m\do\n\do\o\do\p\do\q\do\r\do\s\do\t%
      \do\u\do\v\do\w\do\x\do\y\do\z\do\A\do\B\do\C\do\D%
      \do\E\do\F\do\G\do\H\do\I\do\J\do\K\do\L\do\M\do\N%
      \do\O\do\P\do\Q\do\R\do\S\do\T\do\U\do\V\do\W\do\X%
      \do\Y\do\Z}
\def\UrlDigits{\do\1\do\2\do\3\do\4\do\5\do\6\do\7\do\8\do\9\do\0}
\g@addto@macro{\UrlBreaks}{\UrlOrds}
\g@addto@macro{\UrlBreaks}{\UrlAlphabet}
\g@addto@macro{\UrlBreaks}{\UrlDigits}
\makeatother
\def\x{{\mathbf x}}
\def\L{{\cal L}}

\title{FULLSUBNET: A FULL-BAND AND SUB-BAND FUSION MODEL FOR REAL-TIME SINGLE-CHANNEL SPEECH ENHANCEMENT}

\name{Xiang Hao$^{1,2}$, Xiangdong Su$^{2}$\sthanks{Equal contribution}, Radu Horaud$^3$, and Xiaofei Li$^1$\sthanks{Corresponding author}}

\address{
  $^1$Westlake University $\text{\&}$ Westlake Institute for Advanced Study, Hangzhou, China \\
  $^2$College of Computer Science, Inner Mongolia University, Hohhot, China \\
  $^3$Inria Grenoble Rh\^{o}ne-Alpes, Montbonnot Saint-Martin, France
  }

\begin{document}
\ninept
\maketitle
\begin{abstract}
    This paper proposes a full-band and sub-band fusion model, named as FullSubNet, for single-channel real-time speech enhancement. Full-band and sub-band refer to the models that input full-band and sub-band noisy spectral feature, output full-band and sub-band speech target, respectively. The sub-band model processes each frequency independently. Its input consists of one frequency and several context frequencies. The output is the prediction of the clean speech target for the corresponding frequency. These two types of models have distinct characteristics. The full-band model can capture the global spectral context and the long-distance cross-band dependencies. However, it lacks the ability to modeling signal stationarity and attending the local spectral pattern. The sub-band model is just the opposite. In our proposed FullSubNet, we connect a pure full-band model and a pure sub-band model sequentially and use practical joint training to integrate these two types of models' advantages. We conducted experiments on the DNS challenge (INTERSPEECH 2020) dataset to evaluate the proposed method. Experimental results show that full-band and sub-band information are complementary, and the FullSubNet can effectively integrate them. Besides, the performance of the FullSubNet also exceeds that of the top-ranked methods in the DNS Challenge (INTERSPEECH 2020).
\end{abstract}
\begin{keywords}
    FullSubNet, Full-band and Sub-band Fusion, Sub-band, Speech Enhancement
\end{keywords}
\section{Introduction}
\label{sec:intro}
In recent years, deep learning-based single-channel speech enhancement methods have greatly improved speech enhancement systems' speech quality and intelligibility.
These methods are often trained in a supervised setting and can be divided into time-domain and frequency-domain methods.
The time-domain methods~\cite{rethage2018wavenet, tcnn, Hao2019} use the neural network to map noisy speech waveform to clean speech waveform directly.
The frequency-domain methods~\cite{regression_approch, se_overview, lstm_ss, phase_sensitive_bilstm} typically use the noisy spectral feature (e.g., complex spectrum, magnitude spectrum) as the input of a neural model. Learning target is the spectral feature of clean speech or a certain mask (e.g., Ideal Binary Mask~\cite{ibm_as_goal_wang_2005}, Ideal Ratio Mask~\cite{irm}, complex Ideal Ratio Mask (cIRM)~\cite{cIRM}).
In general, due to the high dimension and the lack of apparent geometric structure for the time domain signal, the frequency domain methods still dominate the vast majority of speech enhancement methods.
In this paper, we focus on real-time single-channel speech enhancement in the frequency domain.

In our previous work~\cite{li_narrow-band_2019}, a sub-band-based method was proposed for single-channel speech enhancement.
Unlike the traditional full-band-based methods, the method performed in a sub-band style:
The input of the model consists of one frequency, together with several context frequencies. The output is a prediction of the clean speech target for the corresponding frequency.
All the frequencies are processed independently.
This method is designed on the following grounds.
\textbf{(1)} It learns the frequency-wise signal stationarity to discriminate between speech and stationary noise. It is known that speech is non-stationary, while many types of noise are relatively stationary. The temporal evolution of frequency-wise STFT magnitude reflects the stationarity, which is the foundation for the conventional noise power estimators~\cite{power_est_1,power_est_2} and speech enhancement methods~\cite{conventional_se_1, conventional_se_2}.
\textbf{(2)} It focuses on the local spectral pattern presented in the current and context frequencies. The local spectral pattern has been proved to be informative for discriminating between speech and other signals.
This method was submitted to the DNS challenge~\cite{dns_challege} in INTERSPEECH 2020 and ranked the fourth place out of the 16 real-time track submissions.

The sub-band model meets the DNS challenge's real-time requirement, and the performance is also very competitive.
However, since it cannot model the global spectral pattern and exploit the long-distance cross-band dependencies. Especially for the sub-band with an extremely low signal-to-noise ratio (SNR), the sub-band model can hardly recover the clean speech, while it will be possible with the help of full-band dependency. On the other hand, the full-band models~\cite{regression_approch, se_overview} are trained to learn the regression between the high-dimensional input and output, lacking a mechanism dedicated to the sub-band information, such as the signal stationarity.

This paper proposes a full-band and sub-band fusion model named FullSubNet to address the above problems. 
Based on plenty of preliminary experiments, the FullSubNet is designed as a series connection of the full-band model and sub-band model. In short, the full-band model's output is input to the sub-band model. 
Through effective joint training, these two models are jointly optimized.
The FullSubNet can capture the global (full-band) context while retaining the ability to model signal stationarity and attend the local spectral pattern. 
Like the sub-band model, the FullSubNet still meets the real-time requirement and can exploit future information within a reasonable latency.
We evaluate the FullSubNet on the DNS challenge (INTERSPEECH 2020) dataset. Experimental results show that the FullSubNet prominently outperforms both the sub-band model~\cite{sub_dns_xiaofeili} and a pure full-band model with a larger amount of parameters with the FullSubNet, which indicates that the sub-band information and the full-band information are complementary. The proposed fusion model is effective for integrating them. Besides, we also compare the performance with the top-ranked methods in the DNS challenge, and the results show that our objective performance measures are better than them.

\begin{figure}[!t]
    \centering
    \centerline{\includegraphics[width=8.5cm]{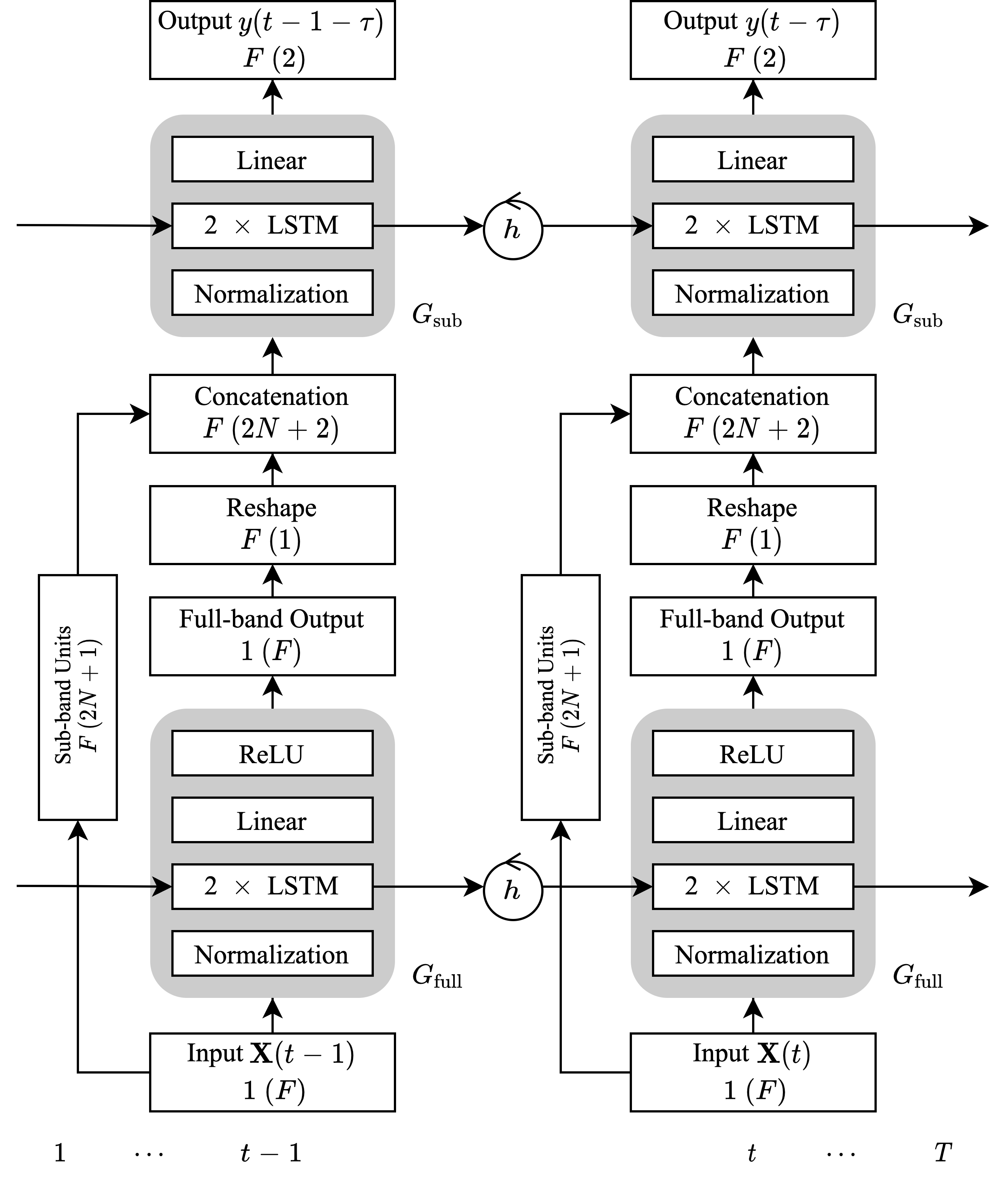}}
    \caption{Diagram of the proposed FullSubNet. The second line in the rectangle describes the dimensions of the data at the current stage, e.g., ``$1~(F)$" represents one $F$-dimensional vector. ``$F~(2N + 1)$" represents $F$ independent ($2N + 1$)-dimensional vectors.}
    \label{fig:workflow}
\end{figure}

\section{Method}
\label{sec:method}

We use the representation of speech signal in the short-time fourier transform (STFT) domain:
\begin{equation}
    X(t,f) = S(t,f) + N(t,f).
\end{equation}
where $X(t,f)$, $S(t,f)$ and $N(t,f)$ respectively represent the complex-valued time-frequency (T-F) bin of noisy speech, noise-free speech (the reverberant image signal received at the microphone) and interference noise at time frame $t$ and frequency bin $f$ with $t = 1, \cdots, T$ and $f = 0, \cdots, F-1$. $T$ and $F$ denote the total number of frames and frequency bins, respectively.

This paper focuses only on the denoising task, and the target is to suppress noise $N(t,f)$ and recover the reverberant speech signal $S(t,f)$.
We propose a full-band and sub-band fusion model to accomplish this task, including a pure full-band model $G_\text{full}$ and a pure sub-band model $G_\text{sub}$. The basic workflow is shown in Fig.~\ref{fig:workflow}.
Next, we will introduce each part in detail.

\subsection{Input}
Previous works~\cite{se_overview,regression_approch,li_narrow-band_2019,sub_dns_xiaofeili} have proved that magnitude spectral feature can provide crucial clues about the global spectral pattern at full-band, while the local spectral pattern and signal stationarity at sub-band.
Therefore, we use the noisy full-band magnitude spectral features
\begin{eqnarray}
    \mathbf{X}(t) = [|X(t, 0)|, \cdots, |X(t, f)| , \cdots, |X(t, F - 1)|]^T \in \mathbb{R}^{F}.
\end{eqnarray} 
We use its sequence
\begin{eqnarray}
    \widetilde{\mathbf{X}} = (\mathbf{X}(1), \cdots, \mathbf{X}(t), \cdots, \mathbf{X}(T))
\end{eqnarray}
as the input of the full-band model $G_\text{full}$.
Then, $G_\text{full}$ can capture the global contextual information and outputs a spectral embedding with the size being the same as $\widetilde{\mathbf{X}}$, which is expected to provide complementary information to the following sub-band model $G_\text{sub}$.

The sub-band model $G_\text{sub}$ predicts the frequency-wise clean-speech target according to the signal stationarity and local spectral mode encoded in the noisy sub-band signal, as well as the full-band model's output.
In detail, we take a time-frequency point $|X(t, f)|$ and its adjacent $2 \times N$ time-frequency points as a sub-band unit.
$N$ is the number of neighbor frequencies considered on each side. For boundary frequencies, with $f - N < 0$ or $f + N > F - 1$, circular Fourier frequencies are used.
We concatenate the sub-band unit and the output of the full-band model, denoted as $G_{\text{full}}(|X(t, f)|)$, as the input of the sub-band model $G_{\text{sub}}$,
\begin{flalign}
    \label{eq:concat}
    \mathbf{x}(t, f) =& [|X(t, f - N)|, \cdots, |X(t, f - 1)|, |X(t, f)| \\ \nonumber
            & |X(t, f + 1)|, \cdots, |X(t, f + N)|, \\\nonumber
            & G_{\text{full}}(|X(t, f)|)]^T \in \mathbb{R}^{2N + 2}.
\end{flalign}
For the frequency $f$, the input sequence of $G_{\text{sub}}$ is
\begin{eqnarray}
    \widetilde{\mathbf{x}}(f) = (\mathbf{x}(1, f), \cdots, \mathbf{x}(t, f), \cdots, \mathbf{x}(T, f)).
\end{eqnarray}
In this sequence, the temporal evolution along with time axis reflect the signal stationarity, which is an efficient cues to discriminate between speech and relatively stationary noise.
The noisy sub-band spectra (composed of $2N+1$ frequencies) and its temporal dynamics provides the local spectral pattern, which can be learned by the dedicated sub-band model. While the signal stationarity cues and the local pattern are actually present in the input of the full-band model $G_{\text{full}}$ as well, however, they are not especially learned by the full-band model $G_{\text{full}}$.  Consequently, the sub-band model $G_{\text{sub}}$ still learns some extra/different information relative to the full-band model $G_{\text{full}}$.  Meanwhile, the output of the full-band model $G_{\text{full}}$ provide some complementary information not seen by the sub-band model $G_{\text{sub}}$. 

Since the full-band spectral feature $\mathbf{X}(t)$ contains $F$ frequencies, we eventually generate $F$ independent input sequences for $G_\text{sub}$ with a dimension of $2N + 2$ for each.

\subsection{Learning target}
\label{sec:learning_target}
There is no doubt that the precise estimation of phase can provide more hearing perception quality improvement, especially in low signal-to-noise ratio (SNR) conditions.
However, the phase is wrapped in $-\pi \sim \pi$ and has chaotic data distribution, which makes it not easy to estimate.
Instead of estimating the phase directly, like the previous works~\cite{li_narrow-band_2019,sub_dns_xiaofeili}, we adopt the complex Ideal Ratio Mask (cIRM) as our model's learning target. 
Follow~\cite{cIRM}, we use hyperbolic tangent to compress cIRM in training and use inverse function to uncompressed mask in inference ($K=10$, $C=0.1$).
We denote cIRM as $ \mathbf{y}(t, f) \in \mathbb{R}^2 $ for one T-F bin.
The sub-band model takes as input sequence $\widetilde{\mathbf{x}}(f)$ for the frequency $f$ and then predicts the cIRM sequence
\begin{eqnarray}
    \widetilde{\mathbf{y}}(f) = (\mathbf{y}(1, f), \cdots, \mathbf{y}(t, f), \cdots, \mathbf{y}(T, f)).
\end{eqnarray}

\subsection{Model architecture}
Fig.~\ref{fig:workflow} shows the architecture of the FullSubNet.
The full-band and sub-band models in the FullSuNet have the same model structure, including two stacked unidirectional LSTM layers and one linear (fully connected) layer.
The LSTM of the full-band model contains 512 hidden units in each layer and uses ReLU as the output layer's activation function.
The full-band model outputs a $F$-dimensional vector at each time step, with one element for each frequency. The sub-band units are then concatenated with this vector frequency by frequency to form $F$ independent input samples for the sub-band model following Equation~\ref{eq:concat}.
According to our previous experiments, the sub-band model is not necessary to be as large as the full-band model, and thus 384 hidden units are used in each layer of LSTM. According to the settings in~\cite{cIRM}, the output layer of the sub-band model does not use activation functions.
It is important to note that all the frequencies share one unique sub-band network (and its parameters).
During training, considering the limited LSTM memory capacity, the input-target sequence pairs are generated with a constant-length sequence.

To make the model easier to optimize, the input sequence must be normalized to equalize the input levels.
For the full-band model, we empirically calculate the mean value $\mu_{\text{full}}$ of the magnitude spectral features on the full-band sequence $\widetilde{\mathbf{X}}$ and normalize the input sequence as $\frac{\widetilde{\mathbf{X}}}{\mu_\text{full}}$.
The sub-band model process the frequencies independently.
For frequency $f$, we caculate the mean value $\mu_\text{sub}(f)$ on the input sequence $\widetilde{\mathbf{x}}(f)$ and normalize the input sequence as $\frac{\widetilde{\mathbf{x}}(f)}{\mu_\text{sub}(f)}$.

In the real-time inference stage, we usually use the cumulative normalization method~\cite{conv_tasnet,DCCRN}, i.e., at each time, the mean value used for normalization is computed using all the available frames. 
However, in the practical real-time speech enhancement system, the speech signal is usually silent initially, which means that the speech signal's beginning part is mostly invalid. 
In this work, to better show the FullSubNet's performance regardless of the normalization problem, we directly use $\mu_\text{full}$ and $\mu_\text{sub}(f)$ computed on the entire test clip to perform normalization during inference.

Same as the method mentioned in~\cite{sub_dns_xiaofeili}, our proposed method supports output delay, which enables the model to explore future information within a reasonably small delay. As shown in the Fig.~\ref{fig:workflow}, to infer $\mathbf{y}(t - \tau )$, the future time steps, i.e. $\mathbf{x}(t - \tau + 1), \cdots, \mathbf{x}(t)$, are provided in the input sequence.

\section{Experimental Setup}

\subsection{Datasets}
We evaluated the FullSubNet on the DNS Challenge (INTERSPEECH 2020) dataset~\cite{dns_challege}.
The clean speech set includes over 500 hours of clips from 2150 speakers. The noise dataset includes over 180 hours of clips from 150 classes.
To make full use of the dataset, we simulate the speech-noise mixture with dynamic mixing during model training.
In detail, before the start of each training epoch, 75\% of the clean speeches are mixed with randomly selected room impulse responses (RIR) from 
\textbf{(1)} the Multichannel Impulse Response Database~\cite{hadad2014multichannel_database} with three reverberation times (T60) 0.16 s, 0.36 s, and 0.61 s. \textbf{(2)} the Reverb Challenge dataset~\cite{kinoshita2016summary_database} with three reverberation times 0.3 s, 0.6 s and 0.7 s.
After that, the speech-noise mixtures are dynamically generated by mixing the clean speech (75\% of them are reverberant) and noise with a random SNR in between -5 and 20 dB. The total data ``seen" by the model is over 5000 hours after ten epochs of training.
The DNS Challenge provides a publicly available test dataset, including two categories of synthetic clips, i.e., without and with reverberations. Each category has 150 noisy clips with SNR levels distributed in between 0 dB to 20 dB.
We use this test dataset for evaluation.

\subsection{Implementation}
The signals are transformed to the STFT domain using a 512-sample (32 ms) Hanning window with a frame step of 256 samples.
We use PyTorch to implement the FullSubNet.
Adam optimizer is used with a learning rate of 0.001.
The sequence length for training is set to $T = 192$ frames (about 3 s).
According to the real-time requirement of the DNS Challenge (INTERSPEECH 2020), we set $\tau = 2$, which exploits two future frames to enhance the current frame, and uses a $16 \times 2 = 32 ms$ look ahead.
As in~\cite{sub_dns_xiaofeili}, we set 15 neighbor frequencies for each side of the input frequency of the sub-band model in the FullSubNet.

\begin{table*}[!t]
    \centering
    \small
    \caption{The performance in terms of WB-PESQ [MOS], NB-PESQ [MOS], STOI [\%], and SI-SDR [dB] on the DNS challege test dataset.}
    \renewcommand\arraystretch{1}
    \setlength{\tabcolsep}{1.3mm}{
        \begin{tabular}{@{}cccccccccccc@{}}
            \toprule
            \multirow{2}{*}{Method}            & \multirow{2}{*}{\begin{tabular}[c]{@{}c@{}}\# Para \\ (M) \end{tabular}} & \multirow{2}{*}{\begin{tabular}[c]{@{}c@{}}Look Ahead \\ (ms) \end{tabular}} & \multirow{2}{*}{Rank} & \multicolumn{4}{c}{With Reverb} & \multicolumn{4}{c}{Without Reverb}                                                                                                         \\
            \cmidrule(lr){5-8}\cmidrule(lr){9-12}
                                               &                                            &                                             &                       & WB-PESQ                         & NB-PESQ                            & STOI           & SI-SDR          & WB-PESQ        & NB-PESQ        & STOI           & SI-SDR          \\
            \midrule
            Noisy                              & -                                          & -                                           & -                     & 1.822                           & 2.753                              & 86.62          & 9.033           & 1.582          & 2.454          & 91.52          & 9.071           \\
            NSNet~\cite{dns_1_nsnet}           & 5.1                                        & 0                                           & -                     & 2.365                           & 3.076                              & 90.43          & 14.721          & 2.145          & 2.873          & 94.47          & 15.613          \\
            DTLN~\cite{DTLN}                   & 1.0                                        &                                             & RT-8                  &                                 & 2.70                               & 84.68          & 10.53           &                & 3.04           & 94.76          & 16.34           \\
            Conv-TasNet~\cite{dns_conv_tasnet} & 5.08                                       & 33                                          & NRT-5                 & 2.750                           &                                    &                &                 & 2.73           &                &                &                 \\
            DCCRN-E~\cite{DCCRN}                 & 3.7                                        & 37.5                                        & RT-1                  &                                 & 3.077                              &                &                 &                & 3.266          &                &                 \\
            PoCoNet~\cite{PoCoNet}             & 50                                         &                                             & NRT-1                 & 2.832                           &                                    &                &                 & 2.748          &                &                &                 \\
            \midrule
            Sub-band Model ~\cite{sub_dns_xiaofeili}  & 1.3                                        & 32                                          & RT-4                  & 2.650                           & 3.274                              & 90.53          & 14.673          & 2.369          & 3.052          & 94.24          & 16.153          \\
            Full-band Model                        & 6.0                                        & 32                                          & -                     & 2.681                           & 3.344                              & 90.64          & 13.580          & 2.731          & 3.256          & 95.71          & 16.190          \\
            FullSubNet                            & 5.6                                        & 32                                          & -                     & \textbf{2.969}                  & \textbf{3.473}                     & \textbf{92.62} & \textbf{15.750} & \textbf{2.777} & \textbf{3.305} & \textbf{96.11} & \textbf{17.290} \\
            \bottomrule
        \end{tabular}
    }
    \label{tab:baselines}
    \vspace{-.6cm}

\end{table*}

\subsection{Baselines}
To testify the full-band and sub-band fusion method's effectiveness, we compare with the following two models, which use the same experimental settings and learning target (cIRM) as the FullSubNet.

\noindent \textbf{$\boldsymbol{\cdot}$ Sub-band model}~\cite{sub_dns_xiaofeili}: The sub-band model has achieved very competitive performance in the DNS-Challenge (the fourth place of the real-time track). To compare performance fairly, like to train the FullSubNet, we use dynamic mixing during training.

\noindent \textbf{$\boldsymbol{\cdot}$ Full-band model}: We construct a pure full-band model, which contains three LSTM layers with 512 hidden units for each layer.  The full-band model's architecture, i.e., a stack of LSTM layers, is actually widely used for speech enhancement, such as in~\cite{se_lstm_noise_asr,lstm_ss}. This model is slightly larger than the proposed fusion model, and thence the comparison would be fair enough.

In addition to these two models, we also compared with the top-ranked methods in the DNS challenge (INTERSPEECH 2020), including NSNet~\cite{dns_1_nsnet}, DTLN~\cite{DTLN}, Conv-TasNet~\cite{dns_conv_tasnet}, DCCRN~\cite{DCCRN} and PoCoNet~\cite{PoCoNet}.

\section{Results}
\subsection{Comparison with the baselines}
In the last three rows of Table~\ref{tab:baselines}, we compare the performance of the sub-band model, the full-band model, and the FullSubNet. ``\# Para'' and ``Look Ahead" in the table respectively represent the parameter amount of the model and the length of used future information. ``With Reverb" means that the noisy speeches in the test dataset have not only noise but also a certain degree of reverberation, which significantly increases the difficulty for speech enhancement. ``Without Reverb" means that the noisy speeches in the test dataset have only noise. For a fair comparison, these three models use the same training target (cIRM), experimental settings, and look ahead.

From the table, we can find that most of the full-band model's evaluation scores are better than the ones of the sub-band model, as the full-band model exploits the wide-band information using a larger network. It is interesting to find that, relative to the full-band model, the sub-band model seems more effective for the ``With Reverb" data, as the superiority of the full-band model for ``With Reverb" is smaller than the one for ``Without Reverb." This indicates that the sub-band model effectively models the reverberation effect by focusing on the temporal evolution of the narrow-band spectrum. The possible reason is that the cross-band dependency of the reverberation effect is actually much lower than the one of signal spectra.

Regarding the FullSubNet: 
\textbf{(1)} Although the sub-band model's performance is already very competitive, after integrating the full-band model (stacked by two LSTM layers and one linear layer), the model performance has been dramatically improved. This improvement shows that the global spectral pattern and the long-distance cross-band dependencies are essential for speech enhancement.
\textbf{(2)} The performance of the FullSubNet also significantly exceeds the full-band model. We must first point out that this improvement does not come from using more parameters. In fact, the FullSubNet (two layers of full-band LSTM plus two layers of sub-band LSTM) has even fewer parameters than the full-band model (three layers of full-band LSTM). After integrating the sub-band model, the FullSubNet inherits the sub-band model's unique ability, namely exploiting signal stationarity and local spectral patterns, and the capability of modeling the reverberation effect.
The apparent superiority of the FullSubNet over the full-band model demonstrates that the information exploited by the sub-band model is indeed not learned by the full-band model, which is complementary to the full-band model. Overall, these results testify that the proposed fusion model successfully integrates the virtues of full-band and sub-band techniques.

\subsection{Comparison with the state-of-the-art methods}

In Table~\ref{tab:baselines}, in addition to showing that the FullSubNet can effectively integrate two complementary models, we also compare its performance with the top-ranked methods in DNS Challenge (INTERSPEECH 2020). 
The ``Rank" column in the table indicates whether to support real-time processing and the challenge's ranking.
e.g., ``RT-8" means the eighth place of the real-time (RT) track. ``NRT-1" means the first place of the non-real-time (NRT) track.

In Table~\ref{tab:baselines}, NSNet is the official baseline method of the DNS challenge, which uses a compact RNN to enhance the noisy short-time speech spectra in a single-frame-in, single-frame-out manner. We use the DNS challenge recipe provided in the asteroid toolkit~\footnote{\url{https://github.com/mpariente/asteroid/tree/master/egs/dns_challenge}} to implement and train NSNet. The training data are generated using the method mentioned in~\cite{sub_dns_xiaofeili}.
In the table, no matter which metric, our proposed method greatly surpasses NSNet with all metrics.

DTLN, Conv-TasNet, DCCRN, and PoCoNet are the top-ranked methods in the DNS challenge's subjective listening test. To ensure the fairness of comparison, we directly quote performance scores from their original papers. The vacant place in the table means that the corresponding score was not reported in the original paper.
DTLN~\cite{DTLN} is capable of real-time processing. It combines the STFT operation and a learned analysis and synthesis basis into a stacked-network with less than one million parameters.
~\cite{dns_conv_tasnet} proposed a low-latency Conv-TasNet. Conv-TasNet~\cite{conv_tasnet} is a widely-used time-domain audio separation network, which has a large computational complexity. Consequently, the low-latency Conv-TasNet does not satisfy the real-time requirement.
DCCRN~\cite{DCCRN} simulates the complex-valued operation inside the convolution recurrent network. It won the first place of the real-time track.
PoCoNet~\cite{PoCoNet} is a convolutional neural network with frequency-positional embeddings employed. Besides, a semi-supervised method is adopted to increase conversational training data by pre-enhancing the noisy datasets. It won the first place of the non-real-time track.
These methods cover a large range of advanced deep learning-based speech enhancement techniques and represent the state-of-the-arts to an extent. 
The original paper of these methods provided the evaluation results on the same test set used in this work but with not all the metrics used in this work. 
It can be seen that the proposed fusion model achieves considerably better objective scores than all of them on this limited dataset.
The performance of PoCoNet is close to ours, but it is a non-real-time model with a much larger network (about 50 M parameters).
The proposed FullSubNet provides a new model dedicated to the full-band/sub-band fusion, which is likely not conflicting with the advanced techniques employed in these state-of-the-art models. Therefore, it worth expecting that speech enhancement capability can be further improved by properly combining them.  

Regarding the computational complexity, the one STFT frame (32 ms) processing time of the proposed model (PyTorch implementation) is $10.32$ ms tested on a virtual quad-core CPU (2.4 GHz) based on Intel Xeon E5-2680 v4, which obviously meets the real-time requirement. 
Later, we will open-source the code and pre-trained models, and show some enhanced audio clips at {\footnotesize\url{https://github.com/haoxiangsnr/FullSubNet}}.

\section{Conclusion}
In this paper, we propose a full-band and sub-band fusion model, named as FullSubNet, for real-time single-channel speech enhancement.
This model is designed to integrate the advantages of the full-band and the sub-band models, that is, it can capture the global (full-band) spectral information and the long-distance cross-band dependencies, meanwhile retaining the ability to modeling signal stationarity and attending the local spectral pattern.
On the DNS challenge (INTERSPEECH 2020) test dataset, we demonstrated that the sub-band information and the full-band information are complementary, and the FullSubNet can effectively integrate them. We also compared the performance with some top-ranked methods for the DNS challenge, and the results show that the FullSubNet outperforms these methods.
\bibliographystyle{IEEEbib}
\bibliography{fusion_model}

\end{document}